\documentclass[10pt, a4paper]{article}

\usepackage[utf8]{inputenc}
\usepackage{authblk}
\usepackage{amsmath}
\usepackage{hyperref} 
\usepackage{amssymb}
\usepackage{graphicx}
\usepackage{lipsum} 

\hypersetup{
    colorlinks=true,
    linkcolor=blue,
    citecolor=blue,
    filecolor=magenta,
    urlcolor=cyan,
    pdftitle={Photonic and phononic modes in acoustoplasmonic toroidal nanopropellers},
    pdfauthor={Autor Uno, Autor Dos, Autor Tres},
    pdfsubject={Asunto del Paper},
    pdfkeywords={Palabras clave del Paper},
    pdfpagemode=FullScreen,
    bookmarks=true,
    pdfborder={0 0 0} 
}

\usepackage[top=1in, bottom=1in, left=1in, right=1in]{geometry}

\title{\textbf{Photonic and phononic modes in acoustoplasmonic toroidal nanopropellers}}

\author[1]{Beatriz Castillo López de Larrinzar}
\author[1]{Jorge M. García}
\author[2]{Norberto Daniel Lanzillotti-Kimura}
\author[1]{Antonio García-Martín\thanks{Corresponding author: a.garcia.martin@csic.es}}

\affil[1]{Instituto de Micro y Nanotecnolog\'ia IMN-CNM, CSIC, CEI UAM+CSIC, Isaac Newton 8, Tres Cantos, 28760 Madrid, Spain}
\affil[2]{Université Paris-Saclay, CNRS, Centre de Nanosciences et de Nanotechnologies,  91120 Palaiseau, France}

\date{}

\begin{document}

\maketitle

\abstract{Non-conventional resonances, both acoustic and photonic, are found in metallic particles with a toroidal nanopropeller geometry that is generated by sweeping a three-lobed 2D-shape along a spiral with twisting angle, $\alpha$. For both optical and acoustic cases, spectral location of resonances experiences a red-shift as a function of $\alpha$. We demonstrate that the optical case can be understood as a natural evolution of resonances as the spiral length of the toroidal nanopropeller increases with $\alpha$, implying a huge helicity dependent absorption cross section. In the case of acoustic response, two red-shifting breathing modes are identified. Additionally, even small $\alpha$ allows the appearance of new low-frequency resonances, whose spectral dispersion depends on a competition between  length of the generative spiral and the pitch of the toroidal nanopropeller. }
\section{Introduction}

The engineering of nanoacoustic and nanophotonic structures is relevant for the conception of novel optomechanical systems, for quantum and sensing technologies, as well as for telecommunications, among others \cite{APL.Perspectives, RevModPhys.OM, Delsing_2019}. The size, shape, and materials of a nanostructure determine simultaneously  its elastic and electromagnetic  behaviors.  The development of nanofabrication techniques over the last thirty years has enabled access to nanostrutures with unprecedented length scales, frequencies, and resolution \cite{Liu2009, NDLK_Metasurfaces, Bragas_Focus, Micropillars_HF, Valentine2008}.  

Objects are considered chiral when their mirror images cannot be superimposed, even after spatial rotations in three dimensions. The importance of chiraly is such that it is critical in life systems since biologically relevant molecules are predominantly chiral (proteins, amino acids, etc.) \cite{drugs}.
Handedness (right vs. left, dextro vs. levo) identifies the two different enantiomers in a chiral structure, and it is the crucial element determining how systems interact with their environment  \cite{Barron_2004}.

Thus, the presence of chirality in a system, unequivocally leads to a geometrical broken symmetry. Good examples of macroscopic systems are  helices, nuts or screws. Particularly interesting is the nut-screw pair. In resemblance to what happens with a screw with the right helicity (and pitch) that nicely couples with the complementary nut, and does not with the other, circularly polarized light interacts with a chiral structure in a different manner depending on the helicity (left or right, levo or dextro) of its circular polarization. In resonant systems, such as metallic ones that support plasmon excitations, the optical interaction with the nanostructure is drastically enhanced. Therefore, the differences in the optical response for levo or dextro light incidence (known as Chiro-Optical Effect - COE) become prominent \cite{Schaeferling2012, IsaMarzan, IsaMarzan2} . Among the plethora of structures to be employed for chiral applications, toroidal propeller-like structures present peculiar properties such as COE reversal \cite{NanoLettPropellers}. 

Macroscopic toroidal propellers are used in aviation and maritime transport to implement less noisy propulsion systems, as they lead to a reduction of tip cavitation and fluid vortices creation since the propeller have no tips. Aquatic toroidal propellers are also improving the fuel efficiency. Although in the aforementioned cases the actual fabrication of these structures by means of classical machining methods is complex, new additive manufacturing techniques may soon charge drastically this scenario.
In the nanoscopic world, advanced nanofabrication tools have enabled the possibility to fabricate these systems by means of scanning electron lithography with an unprecedented resolution and accuracy \cite{Esposito2015, BichiralGiessenTwoPhoton, NanoLettPropellers}.

The development of novel nanostructures possessing simultaneous optical and acoustic resonances is at the core of active fields of research such as optomechanics and nanophononics. In this sense, most of the reported systems rely on relatively simple systems (multilayers, plasmonic antennas, dielectic resonators), \cite{Han, BeaAcustoPlasm, BeaSPIE, Bragas:23, NDLK_PhysRevB.79.035404, NDLK_PhysRevB.83.201103},  where chirality remains elusive.

In this manuscript we develop a complete theoretical study of the optical and acoustic characteristics of a three-lobed toroidal nanopropeller (TnP) made of gold. We do so via the analysis of the scattering and absorption cross sections for the optical response, and calculating the average displacement of the material for an harmonic response after a 5 degree Kelvin thermal uniform expansion for the acoustic one. We demonstrate that the spectral location of the optical plasmon resonances red-shift as a function of the twist angle $\alpha$ of the propeller. Additionally we  observe an interesting interference effect for $\alpha$ in the range of 90 deg. to 270 deg., where the electromagnetic field profile is highly dependent on the spin or helicity of the incident wave. When analyzing the acoustic response we find an even richer and more intriguing behavior.  Through the analysis of the spectra  of the average RMS displacement we demonstrate that the resonance modes supported by the untwisted system also experience a red shift as the twist angle increases. But, at variance with the optical case, the system also allows unexpected new resonant modes at lower energies only present for non-zero twisting angles.

\section{Materials and Methods}
\label{system}

\begin{figure}[ht]
    \centering
    \includegraphics[width=0.6\linewidth]{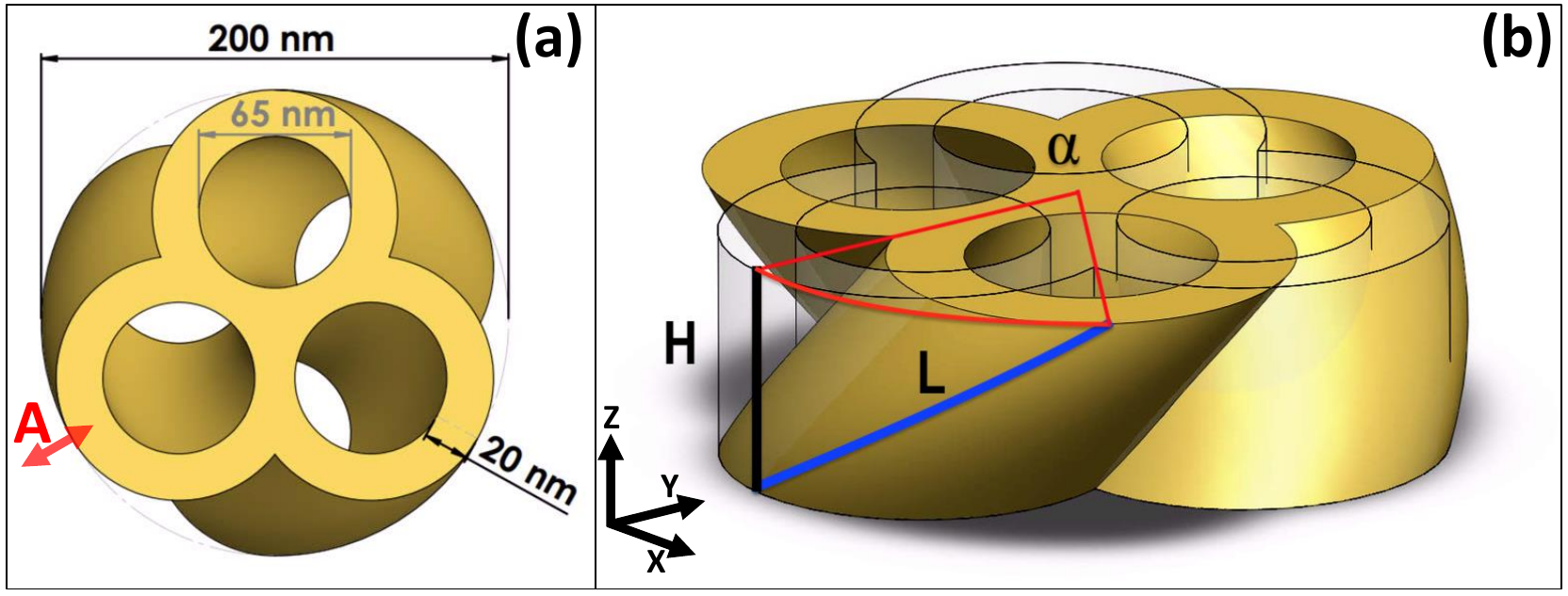}
    \caption{Schematic view of a $\alpha=60$~degrees twisted gold TnP. (a) Top-view depicting the geometrical parameters. The nanostructure is inscribed in a circle of D=200 nm diameter, the diameter of the inner hole is 65 nm and the wall thickness is 20 nm. Label A corresponds to breathing modes resonances (see section~\ref{subsec-acoustic})~(b) The height is H=60 nm, $\alpha$ is the twisting angle. The length of the helical path (blue line) $L$ increases from H (for $\alpha=0$) to $L$ as the twisting angle $\alpha$ increases (see Eq. ~\ref{eq:effective_lenght}).}
    \label{fig:Scheme}
\end{figure}

The geometry of the TnP used in this study is based on a three-lobed structure of identical rings, with an inner diameter of 65 nm and a wall thickness of 20 nm, as presented in Figure~\ref{fig:Scheme}. The centers of the rings are disposed in the vertices of an equilateral triangle, as a result, the whole structure is circumscribed in a circle with a diameter of 200 nm. The twisted TnP is subsequently obtained by vertically sweeping this geometry along a helical path (L) with a varying $\alpha$.  That sweep, together with the vertical dimension (H=60 nm), defines the pitch of the propeller ($P=2\pi H/\alpha $). TnP is a singularly connected structure and with a topological genus 3. 

As mentioned, to simulate the optical and acoustic response, we assume the TnP is made of gold, following the dispersive optical constants given in Ref.~\cite{Johnson_Christy:72} provided in the Lumerical\textregistered~ software, and the thermal expansion, sound speed and Young modulus embedded in the Finite Element Method COMSOL\textregistered. 

Lumerical\textregistered~ software suite has been employed to solve the scattering and absorption cross section. The solutions are based on the Finite Difference Time Domain (FDTD) method.  We use an impinging circularly polarized plane wave with the proper circular polarization along the z axis,
with a normalized amplitude in the whole simulation cell. The cell geometry  (4 $\mu$m × 4 $\mu$m × 7 $\mu$m) ensures that perfectly absorbing boundary conditions do have a negligible effect on the electromagnetic fields obtained. We use a refined mesh in the nanostructures and in the near field region (0.3 $\mu$m × 0.3 $\mu$m × 0.9 $\mu$m) of 2 nm × 2 nm × 2 nm, dx-dy-dz, respectively, growing uniformly up to a maximum of 40 nm out of the near field close to the simulation boundaries, so that convergence (to the best of our numerical capabilities) is attained. 
The total and scattered fields are then collected to give rise to intensity color maps and cross sections.

The Finite Element Method (FEM) within COMSOL\textregistered~ provides a suitable environment to model the acoustic response upon isotropic heat expansion of a metallic nanostructure. In our simulations the toroidal propeller is, as mentioned above, assumed to be made of gold, standing on a layer of silicon dioxide semi-spherical substrate. The material parameters used for  FEM simulation are the ones built-in the materials library of COMSOL\textregistered. The SiO$_2$ substrate is a 800 nm diameter semi-spherical region surrounded by perfectly matching layers (PML) that truncate the physical domain, see e.g. Ref.~\cite{Bragas:23}.

\section{Results}

\subsection{Optical characteristics} \label{optics}
As it can be seen, in a twisted TnP geometry like the one depicted in Figure~\ref{fig:Scheme}b, there is a clear handedness. Therefore, it is expected that the optical response will present an helicity dependence that should, additionally, depend on the twisting angle $\alpha$.  To explore that possibility, the optical response of this structure is interrogated by impinging with a plane wave whose propagation vector is perpendicular in the x-y plane (i.e. in the z direction, from $z=+\infty$), and its polarization is either left-circular (LCP) or right-circular (RCP) as

\begin{equation}
    |LCP\rangle = \frac{|x \rangle + i|y\rangle}{\sqrt{2}}; \quad |RCP\rangle = \frac{|x\rangle - i|y\rangle}{\sqrt{2}},
    \label{}
\end{equation}
\noindent where $|x \rangle$ represents a linearly polarized wave along the x-direction while travelling along the z-direction, and $|y \rangle$ is travelling in the same direction but linearly polarized along the y-direction. 

\begin{figure}[ht]
    \centering
    \includegraphics[width=0.7\linewidth]{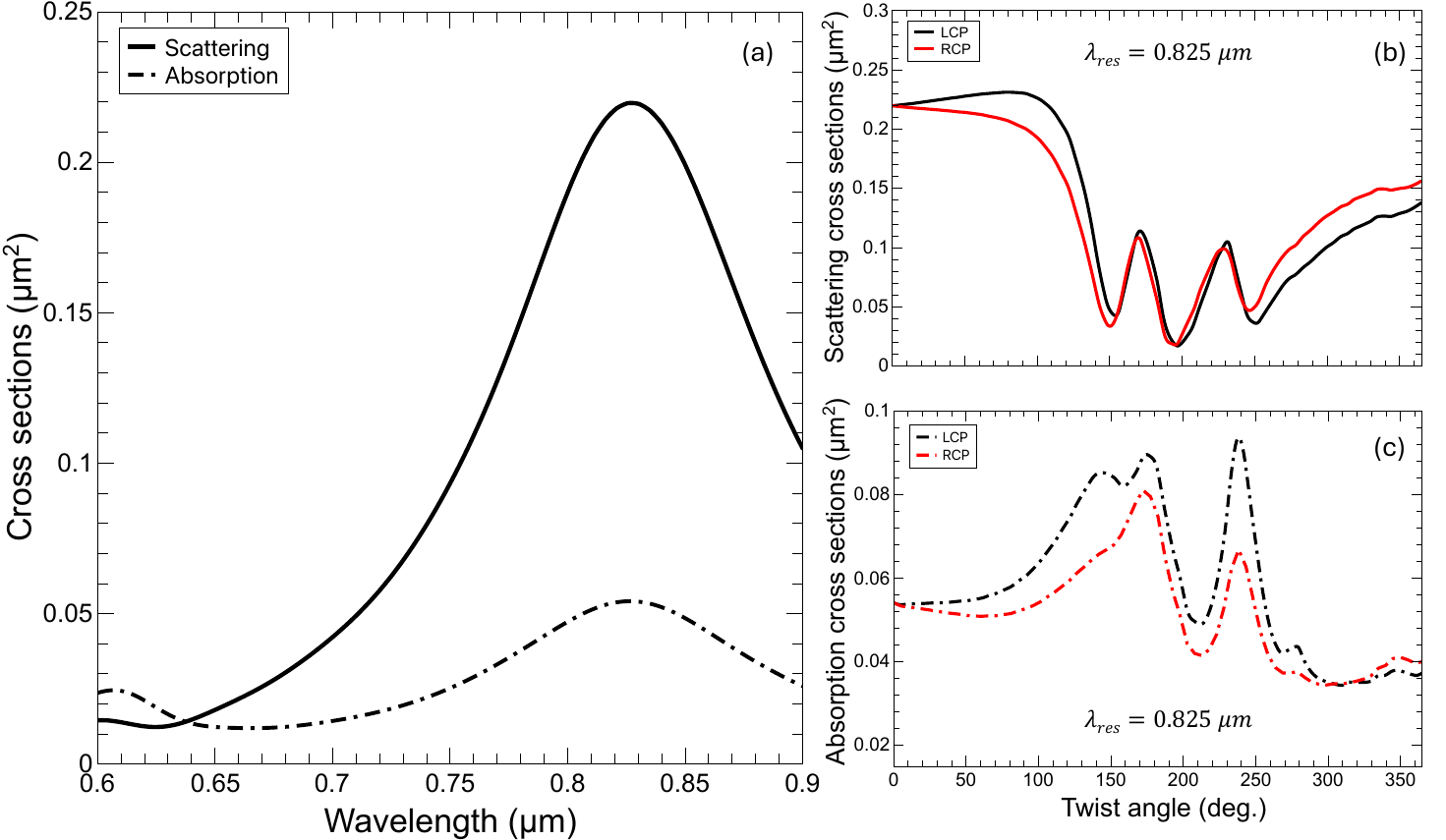}
    \caption{(a) Scattering and absorption cross section for the untwisted TnP in Figure~\ref{fig:Scheme} for LCP and RCP (no differences for this geometry), depicting clearly the main plasmonic resonance at $\lambda_{\text{res}}=0.825~\mu$m, and other weaker at shorter $\sim 0.6 \mu$m wavelengths. (b) Scattering and (c) absorption cross sections for a fixed wavelength, $\lambda_{\text{res}}$, varying the the twisting angle. 
    \label{fig:CS_single} } 
 \end{figure}

Let us first describe the optical response, in the form of scattering and absorption cross sections, for the untwisted nanostructure as a function of the wavelength in the visible range of the optical spectrum. Similarly to a system resembling a metallic gold disc \cite{LionelOneDisk}, the system present a plasmon resonance in the range of 0.8~$\mu$m (see Figure~\ref{fig:CS_single}a). Another smaller resonance is also observed around 0.6~$\mu$m, expected as long as the dielectric constant of Au stays in the metallic range. It is worth noting that for this case there is no difference between LCP and RCP, as foreseen for a structure with no handedness.

In Figure~\ref{fig:CS_single}(b) and (c) we present the scattering and absorption cross sections for the resonance at wavelength $\lambda_{\text{res}}=0.825~\mu$m, as a function of the twisting angle. Noticeably, in this case there are big differences between LCP and RCP incoming waves, signaling the presence of chirality. However, the most prominent feature is the  oscillatory pattern in optical response. In this structure, the only relevant parameter, from the point of view of the optical resonances, that changes with the twisting angle is the length of the pores forming the lateral surface of the propeller. The length of the helix used for generating the TnP (see Figure~\ref{fig:Scheme}(b)) can be expressed as

\begin{equation}
    L = \sqrt{\left (\frac{ \alpha D }{2}\right )^2  + H^2 },
    \label{eq:effective_lenght}
\end{equation}

\noindent where $\alpha$ is in radiands, \textit{D} is the radius of the helix, and \textit{H} the height as mentioned above (see Figure \ref{fig:Scheme}b). Notice that the radius of the helix, optically speaking, will have a different dimension depending on the nature of the resonance, i.e. weather the field is predominantly in the air or in the metal.

\begin{figure}[ht]
    \centering
   \includegraphics[width=0.65\linewidth]{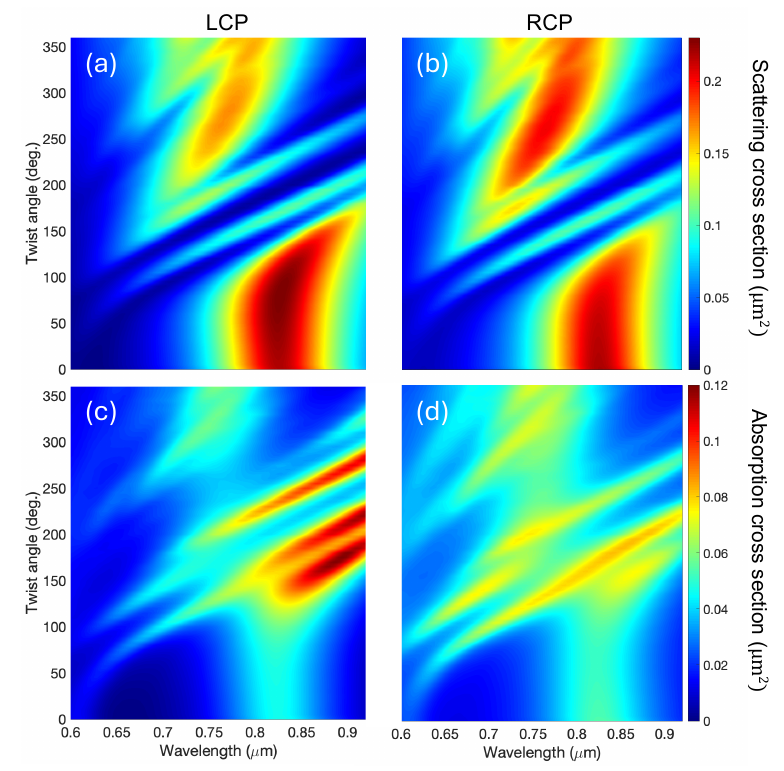}
    \caption{Top: Scattering cross sections for LCP (a) and RCP (b) as a function of the twist angle and of the wavelength. Bottom: Absorption cross sections for LCP (c) and RCP (d) as a function of the twist angle and of the wavelength. As it can be seen, there is an almost linear dependence of the resonances for increasing twist angles. The high helicity dependence, COE, of the absorption is clearly observed in the bottom panels.  }
    \label{fig:CS_angle}
\end{figure}

In order to complete the optical analysis, it is apparent that the complete spectrum of the scattering and absorption cross sections needs to be obtained. This is shown in Figure~\ref{fig:CS_angle}, where we present the scattering and absorption cross sections as a function of the wavelength, for twisting angles bellow 360 degrees. 

In all panels in Figure~\ref{fig:CS_angle} we observe that the resonances at 0.825 $\mu$m and at 0.6$\mu$m, as a function of the twisting angle, behave in the following manner: for $\alpha$ from 0 to c.a. 50 deg. there is very little dispersion, for values from 50 to $\sim$110 deg. the resonances strongly shifts to longer wavelengths, and, interestingly,  for twisting angles $\alpha\gtrsim$120 deg. they disperse linearly towards high wavelengths running parallel to each other.
This happens both for the absorption and the scattering cross sections. Noticeably, for the twisted structure ($\alpha \sim 120$ degrees), the absorption (depicted in (c) and (d)) is over two times bigger than for the untwisted one, but the volume of the system remains essentially the same.
There are obvious differences in the intensity for LCP and RCP waves, pointing out COE, but, apart from the actual value of the cross sections, the overall oscillatory phenomenology is virtually equivalent for both helicities. In fact,  the plot of  Equation ~\ref{eq:effective_lenght}  (see Appendix \ref{Ann:HLength}) shows that it accurately describes the trend followed by the resonance peaks, even in the fact that some may have different slopes (that would depend on a varying effective D). 

Additionally, from a simple inspection of the panels in Figure~\ref{fig:CS_angle} there are obvious signatures of a sizeable COE. This COE is specially relevant in the absorption cross section, where the response of the propeller is enhanced for LCP waves respect to RCP ones, reaching differences larger than 100\%. For the scattering cross section, the differences are not as marked and occur at shorter wavelengths than for the absorption.  

\subsection{Acoustic characteristics}
\label{subsec-acoustic}

Let us now address the characterization of the acoustic response of the TnP. For that end we study the vibrations, characterized by the displacement of each point in the harmonic regime upon an initial isotropic thermal expansion produced by a $\Delta T =5$K. 

\begin{figure}[ht]
    \centering
    \includegraphics[width=0.8\textwidth]{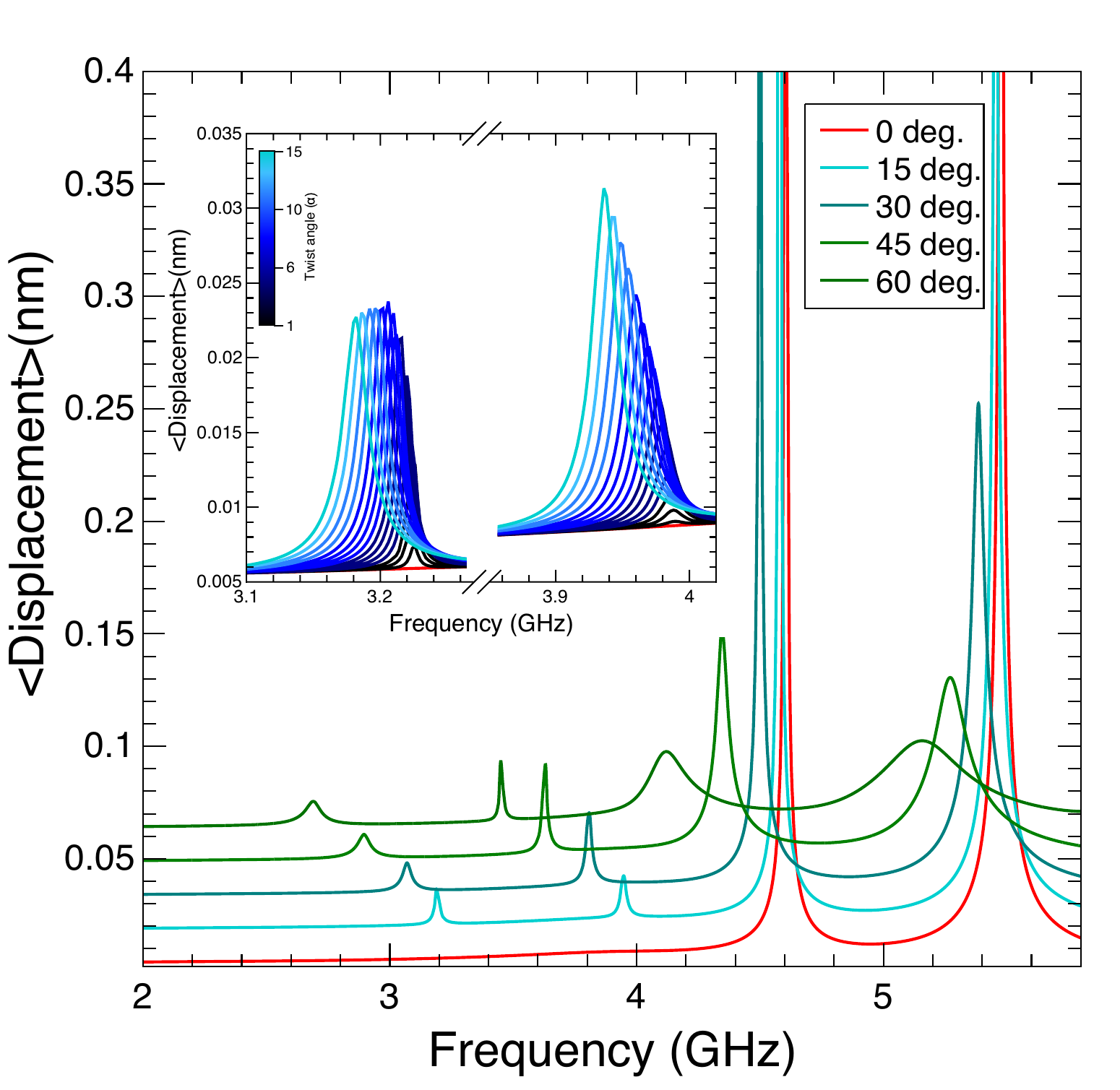}
    \caption{Average RMS displacement of a gold TnP as a function of the acoustic frequency for different twisting angles (0 to 60 degrees). The curves for non-zero twist angle are shifted 0.015~nm vertically for clarity. The peaks represent acoustic phononic modes of the structure. In the inset we present the angular evolution (1 to 15 degrees) of the low frequency modes that appear only for non-zero twist angles.}
    \label{fig:Low_angles_AC}
\end{figure}

In order to find the resonant modes corresponding to the structure, we monitor the average displacement (i.e. the 3D integral of the RMS displacement of each point of the TnP normalized to its volume). The resonant modes appear as distinct peaks as presented in Figure~\ref{fig:Low_angles_AC}. Two natural, low frequency, modes for an untwisted (red curve) TnP appear at 4.6~GHz and 5.5~GHz respectively and correspond to the breathing modes with maximum  (low frequency) and minimum  (high frequency) displacement  at the point marked as A in Figure~\ref{fig:Scheme}. As soon as the structure is slightly twisted (even just one degree) we clearly observe two different phenomena. The first one is that the resonances associated with the breathing modes shift towards lower energies (i.e. the wavelength associated to the resonances increases), in perfect correspondence with the previously presented optical behavior. The second observation, now at variance with the optical case, is that in the 3 to 4~GHz region two new resonances appear that disperse downshift in frequency as the twist angle increases.  
For zero twist angle (red curve in Figure~\ref{fig:Low_angles_AC} and its inset) there are no resonances in this region, but even for a twist of one degree, they appear (black curve inset Figure~\ref{fig:Low_angles_AC}). The evolution with twist angle from $\alpha$=0 to $\alpha$=15 deg. (see inset in Figure~\ref{fig:Low_angles_AC}) shows a further increase in intensity and a downshifting in frequencies. In Figure~\ref{fig:Disp-fields} in Appendix~\ref{Ann:ACModes} we present some snap-shots for the breathing modes at $\alpha=0$~deg. and for the low frequency twist-enabled modes at $\alpha=15$~deg..  

\begin{figure}[ht!]
    \centering
    \includegraphics[width=0.8\linewidth]{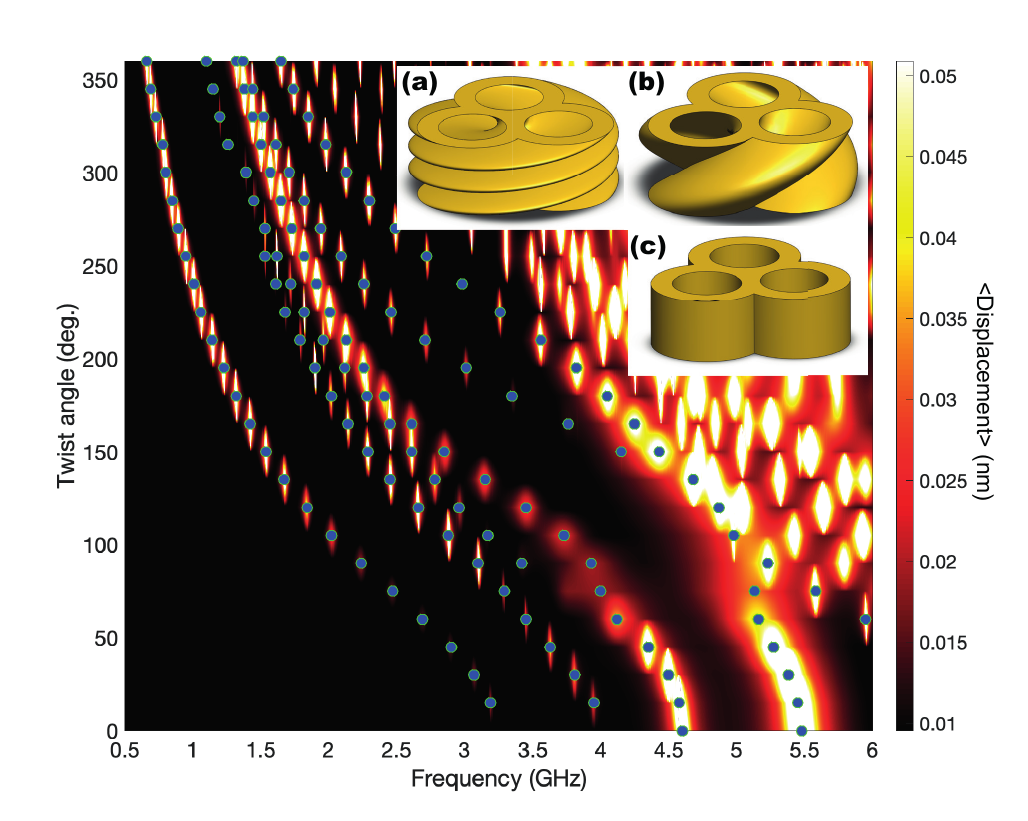}
    \caption{Colormap of the average RMS displacement of a gold TnP as a function of the acoustic frequency for one complete loop of the twisting angle. The intensity is saturated at 0.05~nm for clarity, and the peaks representing the five lowest acoustic phononic modes of the structure are marked with dots. The insets show the geometry for a TnP with a twist angle of 360 degrees (a), 120 degrees (b) and untwisted (c).}
    \label{fig:All_angles_AC}
\end{figure}

Similarly to the optical case, we are presenting in Figure~\ref{fig:All_angles_AC} the whole evolution of the resonance location for a TnP with a complete twist angle (360 deg.). As already indicated in Figure~\ref{fig:Low_angles_AC}, we can see that the lowest energy resonances for the untwisted geometry (inset (c) of Figure~\ref{fig:Low_angles_AC}) appear for 4.6 GHz and 5.5 GHz and evolve towards lower energies as the twist angle increases. One relevant aspect that can be extracted from this figure is that, although the resonances always evolve towards a lower energy value, there is a change in the trend. Initially, the frequency decreases roughly as 1/L as expected to fit with the increase of helix lenght (see Equation ~\ref{eq:effective_lenght}). However, this trend changes at $\alpha\sim 120$~deg.. To shed some light into the change in slope, we refer to the  features depicted in the insets (a) and (b) in Figure~\ref{fig:All_angles_AC}, showing the geometry for twist angles of 360 degrees and 120 degrees respectively. If the structure were made by a single, isolated lobe, at one full loop ($\alpha$=360 deg.) the height of the TnP would match the pitch of the helix, and that would be the first occasion in which one lobe in the base would have another one exactly above. However, our structure is made of three lobes, and Figure~\ref{fig:All_angles_AC}a,  resembles a much more twisted structure, not surprisingly it appears to have three twists. In fact, as seen in Figure~\ref{fig:All_angles_AC}b, it is at one third of a twist ($\alpha = 120$~deg.) when a lobe of the base have another one above. It is then when two effects compete: the resonance trend to decrease in energy due to the growth of L against the resistance due to the upper and lower branches of the TnP.

Eventually, for a very twisted geometry, the resonance should converge to that of a disk shape with a twisted void geometry inside, so compressed that the variation of the angle should have a negligible effect.

\section{Discussion and conclusions}

Understanding the origin of these optical and acoustic modes is essential to actually use them in optomechanical systems, where the helicity of light can be use as a design parameter. Towards that end, in this work we have presented a complete study of the optical and acoustic properties of a three-lobed twisted toroidal nanoscale propeller-like structure. The presence of the twist in the TnP is indicated by a red-shift in both its optical and acoustic resonances. This red-shift occurs because the wavelength couples with the effective length of the generative helix, which varies with the twist angle of the TnP. Consequently, the twist angle acts as a tuning structural parameter that influences the nature, intensity (which depends on helicity), and spectral positions of the optical and acoustic resonances. Additionally, the  twist present in the TnP leads to new low frequency resonant acoustic modes that are not present in the spectrum of an equivalent untwisted structure.  These new acoustic modes appearing at non-zero twisting angles enable a new knob in the engineering of opto-phononic interactions at the nanoscale.


\section*{Appendix A: Length of the helix}
\label{Ann:HLength}

\begin{figure}[ht!]
    \centering
    \includegraphics[width=0.8\linewidth]{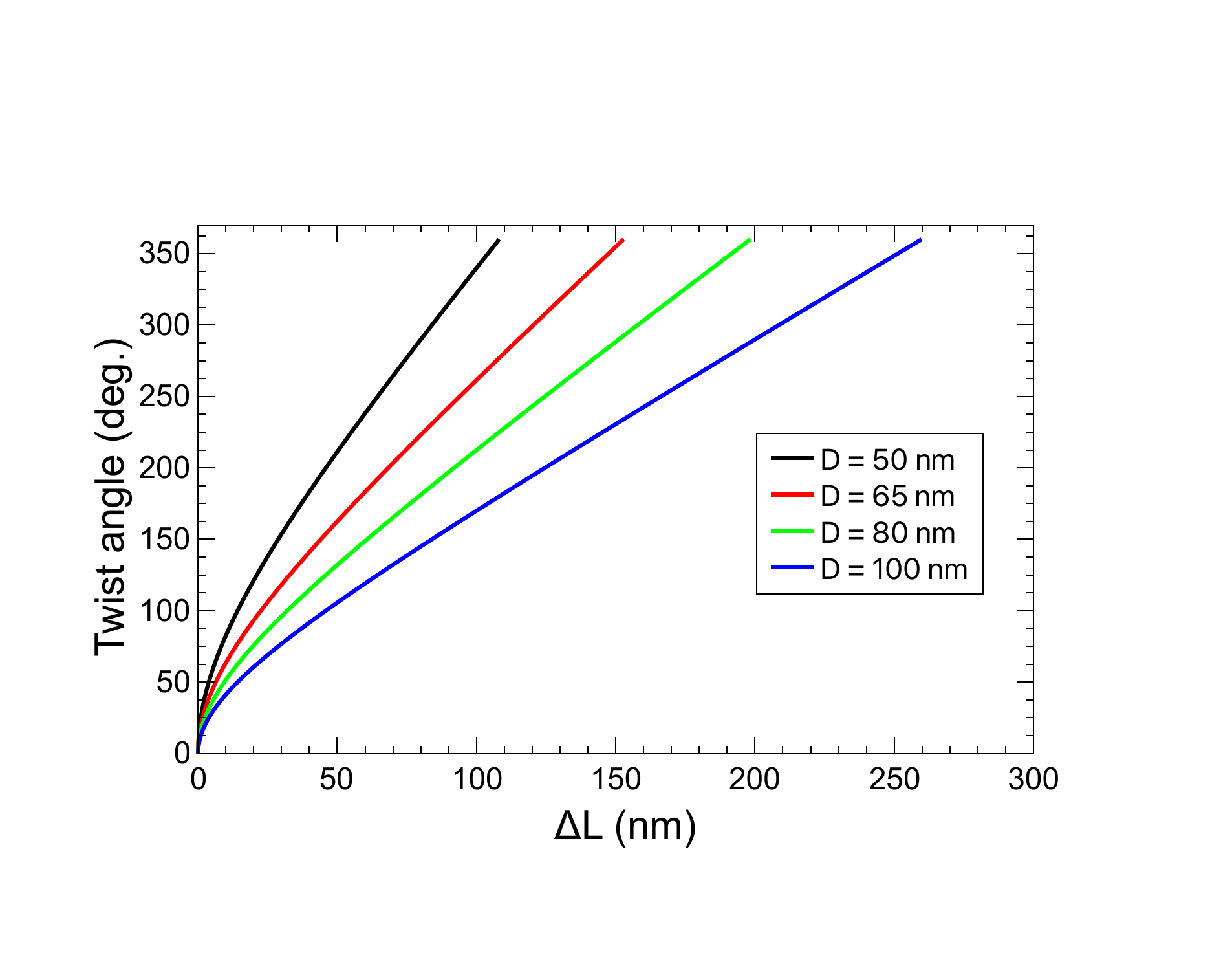}
    \caption{Relationship of the variation of the helix length ($\Delta$L) of the TnP helix as a function of the twisting angle for different values of the parameter \textit{D}.}
    \label{fig:L_vs_alpha}
\end{figure}

In the main text we have argued that the trend followed by the optical and acoustic resonances might be caused by the change in the length of the helical path as the angle increases. This is based on the well-known fact that resonances occur whenever the wavelength is commensurate with a representative length of the system. In our case, the aspect ratio of the TnP changes with $\alpha$ by varying L according to Equation. \ref{eq:effective_lenght}. In Figure~\ref{fig:L_vs_alpha}, we present the behavior of the red-shift that the resonances should experience if the only aspect to be taken into account is the variation of L. As it can be seen, the graph qualitatively describes the physical picture presented in Figures ~\ref{fig:CS_angle} and \ref{fig:All_angles_AC}. Obviously, this has to be taken just as a poor-man phenomenological hand-waving explanation of the actual physics, since there are other aspects playing a key role. Some of them are the dispersive nature of the dielectric permittivity of the materials or the actual path followed by the mode (i.e. where the field or dispersion is maximal), the latter would modify the effective length of the helix, giving rise to different slopes, as seen varying D. 

\appendix
\section*{Appendix B: Displacement profiles}
\label{Ann:ACModes}

\begin{figure}[ht]
    \centering
    \includegraphics[width=0.8\textwidth]{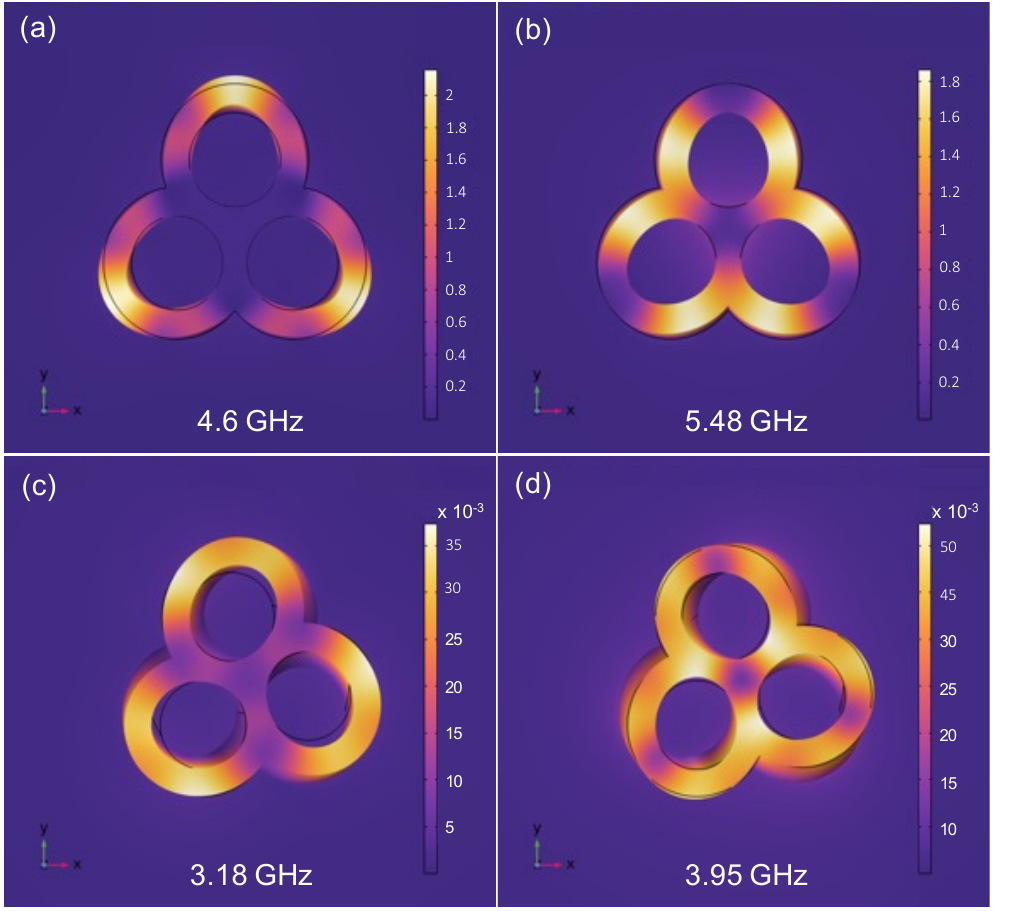}
    \caption{(a) and (b) RMS displacement spatial profile of the two main acoustic modes for the TnP without torsion. (c) and (d) RMS displacement for the twist-allowed modes for a TnP with $\alpha=$15 degrees. The displacement is represented by the color scale and the deformation in the has been scaled for a better view. 
    } 
    \label{fig:Disp-fields}
\end{figure}

In Figure~\ref{fig:Disp-fields}a and b, we present the displacement spatial profiles corresponding to the two lowest energy modes for the untwisted TnP, appearing at 4.6 and 5.48 GHz, respectively (see red lines in Figure~\ref{fig:Low_angles_AC}). We can clearly see that the profiles correspond to two complementary breathing modes: the one with lowest energy corresponds to maximum displacement at the "tip" of each lobe, marked as A in Figure~\ref{fig:Scheme}, whereas the one at highest energy essentially presents a anchor point at that location.

In Figure~\ref{fig:Disp-fields}c and d, we depict the  displacement profiles for a TnP with $\alpha=15$~deg.. In this case the deformation is essentially devoted to exert a torsion in the TnP. The anchor point at Figure~\ref{fig:Disp-fields}c is located at the intersections of the lobes, and at the center of the TnP, allowing more freedom of movement than in the case presented in Figure~\ref{fig:Disp-fields}d. In that case, there is anchor point locate {\em close to} the edge of the lobes, resulting then in a bigger frequency.
\bibliographystyle{ieeetr} 
\bibliography{Preprint} 

\end{document}